\documentclass[usenatbib]{mn2e}
\usepackage[pdfpagemode={UseOutlines},bookmarks,bookmarksopen,colorlinks,linkcolor={black},citecolor={black},urlcolor={black}]{hyperref}
\usepackage{natbibmnfix}
\usepackage{astrojournals}
\usepackage{epsfig}
\usepackage{amssymb}
\usepackage{multirow}
\usepackage{multicol}
\usepackage{color}
\usepackage[varg]{txfonts}

\def \mathbi#1{\textbf{\em #1}}
\def \msun{\rm M_\odot}

\date{}
\pagerange{\pageref{firstpage}--\pageref{lastpage}}
\pubyear{2014}

\begin{document}
\label{firstpage}

\title[A limit on disc viscosity anisotropy]{Accretion disc viscosity: a limit on the anisotropy}
\author[Nixon]{Chris~Nixon\thanks{chris.nixon@jila.colorado.edu}\thanks{Einstein Fellow}
\vspace{0.1in}\\ 
JILA, University of Colorado \& NIST, Boulder CO 80309-0440, USA\\
} 
\maketitle

\begin{abstract}
Observations of warped discs can give insight into the nature of angular momentum transport in accretion discs. Only a few objects are known to show strong periodicity on long timescales, but when such periodicity is present it is often attributed to precession of the accretion disc. The X-ray binary Hercules X-1/HZ Herculis (Her~X-1) is one of the best examples of such periodicity and has been linked to disc precession since it was first observed. By using the current best-fitting models to Her~X-1, which invoke precession driven by radiation warping, I place a constraint on the effective viscosities that act in a warped disc. These effective viscosities almost certainly arise due to turbulence induced by the magneto-rotational instability. The constraints derived here are in agreement with analytical and numerical investigations into the nature of magneto-hydrodynamic disc turbulence, but at odds with some recent global simulations.
\end{abstract}


\section{Introduction}
\label{intro}
Accretion discs \citep{PR1972,Pringle1981,Franketal2002} are found in many astrophysical phenomena, from star and planet formation to the gas accretion on to supermassive black holes which power active galactic nuclei. They also form in stellar mass binaries when, for example, a star fills its Roche lobe and loses mass to a neutron star or black hole companion. Her~X-1 is an X--ray binary in which a neutron star accretes from a donor star. This system displays some stochastic behaviour, but shows a strong 35--day periodicity \citep{Tananbaumetal1972} linked to precession of the accretion disc \citep{Katz1973,Roberts1974,Petterson1975,GB1976}. The precession is plausibly attributed to radiation warping of the disc (\citealt{Pringle1996}; \citealt{Pringle1997}; \citealt{WP1999}; \citealt{OD2001}). 

Accretion discs evolve due to the action of a turbulent viscosity. In most cases this is driven by the magneto-rotational instability \citep[MRI;][]{BH1991}. In some cases, e.g. massive protostellar or AGN discs, the disc self-gravity can also induce turbulence alongside the MRI, but in X-ray binaries the disc mass is small enough that this does not contribute \citep[e.g. eq. 5.51 of][]{Franketal2002}. In the study of warped discs it is often assumed that the effective viscosity arising from the disc turbulence can be treated as isotropic. This means that azimuthal and vertical shear are damped at the same average rates. This effective viscosity is usually modelled as a \cite{SS1973} $\alpha$ viscosity.

Warped discs are additionally complicated by the presence of a radial pressure gradient. This occurs because the midplanes (and therefore regions of highest pressure) are misaligned if two neighbouring regions of the disc are misaligned. Therefore around each ring ($2\pi$ in azimuth) there is an oscillating pressure force \citep{PP1983}. In the limit that $\alpha \ll H/R$ this results in a pressure warp wave being launched radially through the disc \citep{PL1995,PT1995,LO2000,Lubowetal2002}. \cite{IP2008} explore radiation warping dynamics for different disc models, e.g. wavelike discs. If instead, $\alpha \gg H/R$ then the effect is strongly damped locally and the disc warp diffuses. X-ray binary discs are usually in the diffusive scenario as $\alpha \sim 0.1$ \citep{Kingetal2007} and $H/R \sim 0.01$ \citep[eq 5.50][]{Franketal2002}.

In the diffusive warp propagation case, \cite{PP1983} provided the first self-consistent evolution equation for a warped disc, assuming an isotropic effective $\alpha$. Their discovery of the oscillating radial pressure gradient showed that the effective viscosity governing the radial communication of the component of angular momentum parallel to the local orbital plane ($\nu_2 = \alpha_2 c_{\rm s}H$) does not have the same magnitude as the effective viscosity governing the radial communication of the component of angular momentum perpendicular to the plane of the disc ($\nu_1 = \alpha_1 c_{\rm s}H$). In fact, while $\nu_1 \propto \alpha$, $\nu_2 \propto 1/\alpha$.

The reason for this inverse dependence on the turbulent viscosity is the oscillating pressure gradient -- the induced radial shearing motions are damped at a rate given by $\alpha$ and therefore communication is restricted for larger values of $\alpha$. A simple theory of this relation is given by Section 4.1 of \cite{LP2007}, but the full derivation, in the linear and nonlinear cases respectively, is given by \cite{PP1983} and \cite{Ogilvie1999}.

After the initial warped disc equations developed by \cite{PP1983}, \cite{Pringle1992} derived an evolution equation for a warped disc based purely on conservation laws, which describes the evolution of a disc with an arbitrary shape once the effective viscosities $\nu_1$ and $\nu_2$ have been specified. Then, \cite{Ogilvie1999,Ogilvie2000} derived an evolution equation from first principles assuming little more than an isotropic effective $\alpha$. Ogilvie's equations are fully nonlinear and largely confirm the equations derived by \cite{Pringle1992}, but with two important differences: 1) an additional internal torque that makes tilted neighbouring rings precess, 2) the effective viscosity coefficients $\alpha_1$ and $\alpha_2$ are functions of both $\alpha$ and the local disc warp amplitude. The work of \cite{Ogilvie1999,Ogilvie2000} gives a complete theory of the evolution of warped discs for an isotropic $\alpha$ in the diffusive case. For the wavelike case there is no full nonlinear theory \citep[see e.g.][]{Ogilvie2006}, but the linearised equations can be found in \cite{PL1995} and \cite{LO2000}.

Almost all work on warped discs assumes that the effective viscosity arising from MHD turbulence is isotropic. \cite{Pringle1992} noted that the viscosity may well be significantly anisotropic due to the different types of shear present in a warped disc. Azimuthal shear is secular where gas particles drift further apart, whereas vertical shear is oscillatory and thus should induce less dissipation. However, if the velocity spectrum of the turbulence is predominantly on scales $< H$ \citep[e.g. Fig.~14 of ][]{Simonetal2012} then it is likely to act similarly in each direction. Theoretical investigations into the form of MHD disc turbulence are inconclusive, and so it is useful to examine what can be learnt from observations. \cite{Kingetal2013} argue that observations of warps can constrain the angular momentum transport in accretion discs, and in particular $\nu_2$. This builds on \cite{Kingetal2007} who used observations of dwarf novae and soft X-ray transients to constrain the planar disc viscosity $\nu_1$.

In this work I explore the best-fitting models for the observed periodicity in Her~X-1, which has a disc that is tilted and forced to precess by the radiation warping instability, to see what constraints can be provided on the internal communication of angular momentum in a warped disc.

\section{Warped disc viscosity}
The definitions and meanings attached to ``viscosity'' can be quite complex in the case of a warped disc, so I review the relevant equations here. The evolution equation for the angular momentum vector, $\mathbi{L} = \Sigma R^2\Omega \mathbi{l}$ (where $R$ is the radial coordinate, $\Sigma$ is the disc surface density, $\Omega$ is the disc rotation law and $\mathbi{l}= (\cos\gamma\sin\beta,\sin\gamma\sin\beta,\cos\beta)$ is the unit tilt vector for tilt and twist angles $\beta$ and $\gamma$ respectively), in a warped disc \citep{Pringle1992,Ogilvie1999} is
\begin{eqnarray}
\label{dLdt}
 \frac{\partial \mathbi{L}}{\partial t} & = & ~~~\frac{1}{R}
 \frac{\partial }{\partial R} \left\{ \frac{\left(\partial / \partial
   R \right) \left[\nu_{1}\Sigma R^{3}\left(-\Omega^{'} \right)
     \right] }{\Sigma \left( \partial / \partial R \right) \left(R^{2}
   \Omega \right)} \mathbi{L}\right\} \\ \nonumber & &
 +{}\frac{1}{R}\frac{\partial}{\partial R}\left[\frac{1}{2} \nu_{2}R
   \left| \mathbi{L} \right|\frac{\partial \mathbi{l}}{\partial
     R} \right] \\ \nonumber & &
 +{}\frac{1}{R}\frac{\partial}{\partial R}
 \left\{\left[\frac{\frac{1}{2}\nu_{2}R^{3}\Omega \left|\partial
     \mathbi{l} / \partial R \right| ^{2}}{\left( \partial /
     \partial R \right) \left( R^{2} \Omega \right)} +
   \nu_{1}\left(\frac{R \Omega^{'}}{\Omega} \right) \right]
 \mathbi{L}\right\}\\ \nonumber & &
 +{}\frac{1}{R}\frac{\partial}{\partial R}\left(\nu_3 \Sigma R^3 \Omega \mathbi{l}\times\frac{\partial \mathbi{l}}{\partial R}\right)\\ \nonumber & &
 +{}\frac{1}{2\pi R}\frac{{\rm d} \mathbi{G}}{{\rm d} R}\,.
\end{eqnarray}
The first and second terms govern the diffusion of disc surface density and tilt respectively and the third term is an advective term. The coefficients of the torques communicating angular momentum in the disc, $\nu_1$ and $\nu_2$, are the effective viscosities. In reality these result from MHD turbulence and hydrodynamics, but their effects are called viscosities, following a long-established usage in both the astrophysical literature and the fluid dynamics literature. The second last term is the internal precession torque that is present in the analysis of \cite{PP1983}\footnote{The imaginary part of the quantity $A$ defined in their Section 4.2.}, but first put into this form by \cite{Ogilvie1999}. In this term $\nu_3 = \alpha_3 c_{\rm s}H$ (where $\alpha_3 \approx 3/8$; \citealt{Ogilvie1999}), but it is misleading to call this a `viscosity' as it does not lead to diffusion, but instead causes a dispersive wavelike propagation of the warp \citep{Ogilvie1999}. As this term does not significantly affect the dynamics of diffusive warped discs, I do not discuss it further here. The final term corresponds to the external torque on the disc, in this case due to radiation. \cite{Pringle1996,Pringle1997} gives
\begin{equation}
\frac{{\rm d}\mathbi{G}}{{\rm d}R} = -\frac{L_\star}{6\pi c}\oint {\rm d}\phi \frac{A\mathbi{s}_\phi}{(1+A^2)^{1/2}}\,,
\end{equation}
where $L_\star$ is the luminosity of the central source, $c$ is the speed of light, the integral is taken only over illuminated parts of the disc, and $A$ and $\mathbi{s}_\phi$ are given by 
\begin{equation}
A = R \frac{\partial \beta}{\partial R}\sin\phi - R\frac{\partial\gamma}{\partial R}\cos\phi\sin\beta
\end{equation}
and 
\begin{eqnarray}
\mathbi{s}_\phi & = & (\cos\phi\cos\gamma\cos\beta-\sin\phi\sin\gamma, \\ \nonumber & &
 {} \cos\phi\sin\gamma\cos\beta+\sin\phi\cos\gamma, \\ \nonumber & &
 -{} \cos\phi\sin\beta)\,.
\end{eqnarray}

It can be seen from (\ref{dLdt}) that when there are no external torques on the disc (${\rm d}\mathbi{G}/{\rm d}R = \mathbi{0}$), the only steady disc shape is a flat disc. This occurs because the vectors $\mathbi{l}$, $\partial\mathbi{l}/\partial R$ and $\mathbi{l}\times\partial\mathbi{l}/\partial R$ form an orthogonal set. So the independent terms in (\ref{dLdt}) cannot balance each other, and so in the presence of dissipation both the second and fourth terms must decrease to zero, implying that $\left|\partial\mathbi{l}/\partial R\right| = 0$, i.e. a flat disc.

Both effective viscosities, $\nu_1$ and $\nu_2$, can be written in the Shakura--Sunyaev form
\begin{equation}
\label{effvisc}
\nu_1 = \alpha_1 c_{\rm s} H; ~~~ \nu_2 = \alpha_2 c_{\rm s} H\,,
\end{equation}
where $\alpha_1$ and $\alpha_2$ are the effective viscosity coefficients. It is often assumed that an ``isotropic viscosity'' is equivalent to taking these coefficients as equal, but this is not the case. Initial investigations into warped discs \citep[e.g.][]{BP1975,Hatchettetal1981} made this assumption, but it was shown to be inconsistent by \cite{PP1983}. This is because of the internal oscillating pressure gradient described above. Instead, if we assume that $\alpha$ acts isotropically in the disc, damping any type of shear at the same rate, then \cite{PP1983} showed that
\begin{equation}
\label{visccoef}
\alpha_1 = \alpha;~~~\alpha_2 = \frac{1}{2\alpha}\,.
\end{equation}
This is what characterises an isotropic viscosity. Further, \cite{PP1983} give higher order terms in these coefficients and also \cite{Ogilvie1999} extends this to the nonlinear warp regime. See Appendix~\ref{coeffs} for a discussion.

Now, if the disc viscosity is not isotropic then the notation becomes even more complex. We must now split (the previously assumed isotropic) $\alpha$ into two components $\alpha_{\rm h}$ and $\alpha_{\rm v}$, which govern the damping of azimuthal\footnote{The origin of this notation is from shearing box calculations which discuss ``horizontal'' and vertical shear. In a full 3D disc, this corresponds to ``azimuthal'' and vertical shear. However, we are left with $\alpha_{\rm h}$ for azimuthal shear.} and vertical shear respectively. Formally $\alpha_{\rm h}$ corresponds to the horizontal-horizontal components of the rate-of-strain tensor, while $\alpha_{\rm v}$ corresponds to the horizontal-vertical components of the rate-of-strain tensor \citep{Torkelssonetal2000}. When $\alpha_{\rm h} = \alpha_{\rm v} = \alpha$ (the isotropic case) we have effective viscosities given by (\ref{effvisc}) and ({\ref{visccoef}}), but when this is not the case we have
\begin{equation}
\label{anisoalphas}
\alpha_1 = \alpha_{\rm h};~~~\alpha_2 = \frac{1}{2\alpha_{\rm v}}\,.
\end{equation}

In principle these damping rates ($\alpha_{\rm h}$ and $\alpha_{\rm v}$) can be calculated from MHD turbulence calculations and compared to determine if the effective $\alpha$ is isotropic or whether the azimuthal and vertical shear are not damped at the same rates. The initial investigation into this was the shearing box calculations of \cite{Torkelssonetal2000} that showed $\alpha_{\rm h} \approx \alpha_{\rm v}$ in good agreement with an isotropic $\alpha$ viscosity. This was further confirmed by the analytical treatment of MHD turbulence by \cite{Ogilvie2003}.

In any case, it appears sensible to use observations to provide any possible insight into the angular momentum transport in warped discs. In the next section I derive a constraint on the ratio of the vertical and azimuthal effective viscosities, by considering the best-fitting models to the tilt and precession in the X-ray binary system Her~X-1.

\section{Anisotropy constraint}
\label{constraint}
We consider the disc system in Hercules X-1/HZ Herculis (Her~X-1). This is a semi-detached binary system with an F star of mass $\approx 2\msun$ which transfers mass to a neutron star companion of mass $\approx 1\msun$. The binary period is $1.7$~days and the long period associated with precession is $35$~days. Using appropriate parameters for the system, \cite{Kingetal2013} estimate the alignment timescale of the disc with the binary to be of order 10~days. Therefore, as the system has been observed to show this periodicity for $>40$~years, it is clear that some physical process must be maintaining the tilt.

Radiation warping \citep{Petterson1977a,Petterson1977b,Pringle1996,Pringle1997} can induce and maintain the tilt in such systems. This was first suggested for Her~X-1 by \cite{Petterson1977c} and further models have been developed by \cite{WP1999} and \cite{OD2001}. Radiation warping is also successful in reproducing the long period properties of a number of other binary systems, as discussed in \cite{WP1999} and \cite{OD2001}. The models have some unique and appealing properties consistent with a variety of observed effects, which I list here. The radiation warping mechanism is one of only a small number of physical effects that can produce the disc tilt, and once the tilt is induced it also drives precession at a rate consistent with observations -- in comparison, precession forced by the companion's gravity does not reproduce the observed precession rates \citep{WP1999} and does not account for the disc tilt as the disc would align on a short timescale \citep{Kingetal2013}. The precession induced by radiation warping can be prograde or retrograde and therefore account for the unusual prograde precession observed in systems such as Cyg X-2 \citep[e.g.][]{MB1997,Brocksoppetal1999}. If the central source luminosity is too high, the warped disc structure can be non-stationary with oscillating tilt - this may be applicable to many of the long X-ray periods which are not stable in amplitude or period \citep{WP1999}. Another compelling feature of radiation warping is that the inner disc may tilt through more than $90^\circ$, allowing it to accrete on to the central object in a counterrotating manner -- this has successfully explained the torque reversals seen in some neutron star X-ray binaries \citep{vanKerkwijketal1998}. These attributes of the radiation warping model form a compelling case for it being responsible for the long periods in a number of systems such as Her~X-1.

Using the models for Her~X-1 provided by \cite{WP1999} and \cite{OD2001} we can constrain the effective viscosity (angular momentum communication) in the disc through the appropriate instability condition, as the disc must be unstable to radiation warping (but only just so) to undergo repeated precession \citep{OD2001}. \cite{Pringle1996} gives the criterion for a disc to be unstable to radiation warping as
\begin{equation}
\frac{R}{R_{\rm g}} \gtrsim \frac{16\pi^2 \eta^2}{\epsilon^2}\,,
\end{equation}
where $R$ is the disc radius, $R_{\rm g}=GM_1/c^2$ is the Schwarzchild radius of the accretor, $\eta = \nu_2/\nu_1$ is the ratio of vertical over azimuthal effective viscosities and $\epsilon$ is the accretion efficiency. \cite{OD2001} write this criterion in terms of the binary orbit as
\begin{equation}
\frac{R_{\rm b}}{R_{\rm g}} \gtrsim \frac{16\pi^2\eta^2}{\epsilon^2}\frac{R_{\rm b}}{R}\,.
\end{equation}
For Her~X-1, $R_{\rm b}/R_{\rm g} = 3.1 \times 10^6$, and the disc is expected to truncate at $R\approx 0.3 R_{\rm b}$ for this near--equal mass binary with $M_0/M_1 = 1.56$, where $M_0$ is the companion mass \citep{OD2001}. Choosing $R=0.3R_{\rm b}$ is a conservative limit as the instability must act at some radius in the disc $R \lesssim 0.3R_{\rm b}$. For these numbers the criterion becomes
\begin{equation}
\frac{\eta}{\epsilon} \lesssim 75\,.
\end{equation}
Assuming a typical accretion efficiency $\epsilon=0.1$ we now have a simple constraint on the ratio of effective viscosities. For radiation warping to drive the precession observed in Her~X-1, we require
\begin{equation}
\label{crit}
\frac{\nu_2}{\nu_1} \lesssim 7.5\,.
\end{equation}

In the usual, isotropic, picture of accretion disc viscosity (\citealt{SS1973}; \citealt{PP1983}; \citealt{Ogilvie1999}; \citealt{Ogilvie2000}; \citealt{Torkelssonetal2000}; \citealt{Ogilvie2003}; \citealt{Kingetal2007}; \citealt{Kingetal2013}) and in the limit of small tilt, the viscosity ratio can be written as $\eta \approx 1/2\alpha^2$ (\citealt{PP1983}; see \citealt{Ogilvie1999} for the nonlinear corrections due to warp amplitude). In this case the limit is $\alpha \gtrsim 0.25$, consistent with estimates by \cite{WP1999}, \cite{OD2001} and \cite{Kingetal2013}. This limit is also consistent with the suggestion that Her~X-1 is close to the stability limit \citep{OD2001}, as $\alpha \approx 0.25$ \citep{Kingetal2007}\footnote{Existing MHD simulations produce $\alpha$ values that are lower than inferred for Her X-1 and other similar systems. Shearing box MHD simulations find $\alpha \sim 0.01$ (assuming no net vertical magnetic flux). For example \cite{Simonetal2012} find a few $\times~0.01$. Global MHD simulations suggest numbers slightly higher, e.g. Fig. 1 of \cite{Parkin2014} shows an $\alpha$ of $\sim 0.05$. However, there is currently no accepted reason for this discrepancy with observations \citep[see e.g.][]{Kingetal2007}. The observations appear to be correctly interpreted \citep[e.g.][]{KL2012} and the simulations appear to be converged (e.g. \citealt{Simonetal2012}; but see also \citealt{Bodoetal2014}). So the discrepancy seems real, and is therefore probably driven by either a missing piece of input physics or an environmental effect.}. In other words, an isotropic (Navier--Stokes) viscosity is perfectly consistent with the idea that the precessing disc in Her~X-1 arises from radiation warping \citep[see also][]{Kingetal2013}.

If we now instead {\it assume} that the effective viscosity is anisotropic, we have from (\ref{anisoalphas})
\begin{equation}
\frac{\nu_2}{\nu_1} = \frac{\alpha_2}{\alpha_1} = \frac{1}{2\alpha_{\rm h}\alpha_{\rm v}}\,,
\end{equation}
\citep{Torkelssonetal2000}. Therefore the criterion for the radiation warping instability to act is 
\begin{equation}
\label{anisocrit}
\frac{1}{2\alpha_{\rm h}\alpha_{\rm v}} \lesssim 7.5\,.
\end{equation}
If we again take $\alpha_{\rm h} (=\alpha) = 0.25$, then this becomes $\alpha_2 \lesssim 2$, or equivalently $\alpha_{\rm v} > 0.25$. Thus, for a significantly anisotropic viscosity, the action of radiation warping in Her~X-1 places a limit on $\alpha_2$.

There are higher-order corrections to this result, from both the limit of large $\alpha$ \citep{PP1983,KP1985} and the disc rotation law, which is modified by the binary companion. I discuss and derive the modification to this criterion due to both of these effects in Appendices~\ref{coeffs} \& \ref{nonkep}. I find that for the parameters in this study the corrections are relatively small, with less than a factor of two in the constraint. However, I note that for other systems these corrections can be large.

Recently \cite{Sorathiaetal2013b} and \cite{MTetal2014} have argued that the effective viscosities in a warped disc arising from MHD turbulence are significantly anisotropic. For example, \cite{Sorathiaetal2013b} perform a single MHD simulation of a tilted accretion disc and report that $\alpha_{\rm v} \approx 3\times 10^{-5} \ne \alpha_{\rm h}$. \cite{Sorathiaetal2013b} do not report $\alpha_{\rm h}$, but similar simulations yield values $\simeq 0.01$. Therefore the effective viscosity in Sorathia et al.'s simulation is significantly {\it stronger} than isotropy would predict. In their simulation this produces rapid disc alignment, on approximately the (artificially shortened) precession time. Such disc behaviour is not consistent with constraint (\ref{crit}) derived above from the precession observed in Her~X-1. However, it is not clear that these simulations adequately resolve the MRI process \citep[see e.g. Section 2.2 of][]{MTetal2014}. Also \cite{Sorathiaetal2013b} employ a precession timescale (almost as fast as dynamical) and disc parameters ($H/R \gg \alpha$) that puts them into the highly nonlinear wavelike warps regime where there is no full nonlinear theory (even for an isotropic viscosity). Thus it is not clear how to interpret these results. Further, plots of components of the stress tensor as functions of position angles, as found in \cite{Sorathiaetal2013b} and \cite{MTetal2014} are somewhat meaningless when comparing to an $\alpha$ theory, particularly whether the stress is positive or negative, as it is the {\it average} rate of damping that is given by $\alpha$, and not the instantaneous local stress on scales $\ll H\times H$. Thus it appears plausible that simply the wrong quantities are being measured from the simulations. On this point I finally note that \cite{Zhuravlevetal2014} report good agreement (in the wavelike warps case) between a semi-analytical approach based on an isotropic $\alpha$ and the MHD simulations of \cite{MTetal2014}. They also report ``minimal'' differences between an isotropic and anisotropic viscosity (see their Section~4.1). This presumably occurs because the discs are wavelike and thus the short term dynamics are governed by pressure, rather than viscosity.

Other investigations into the nature of MHD turbulence produce results which are consistent with the criterion derived here. \cite{Torkelssonetal2000} measure $\alpha_{\rm h}$ and $\alpha_{\rm v}$ from shearing box calculations and conclude they are both $\approx 0.01$, in agreement with the isotropic case. \cite{Ogilvie2003} followed this up with an analytical investigation and found the same result. I note that the approach of \cite{Torkelssonetal2000} allowed a direct measure of the values of $\alpha_{\rm h}$ and $\alpha_{\rm v}$, thus offering a clean answer, without the need to disentangle the effects of global simulations with more complex physics.

\section{Conclusions}
I have derived a constraint on the anisotropy of the effective viscosities in warped accretion discs. This constraint is found by considering the best-fitting models for the periodicity observed in Her~X-1. These models use the radiation warping mechanism of \cite{Pringle1996}. The constraint is satisfied by the widely used effective viscosities that result from a (near) isotropic $\alpha$, but is in conflict with significantly anisotropic viscosities where the component communicating misaligned angular momentum is significantly stronger than an isotropic model would imply. Recent MHD simulations have suggested the effective viscosity is significantly anisotropic in this manner, but this would imply that radiation warping in Her~X-1 is prohibited. Many other investigations into the nature of MHD turbulence have concluded that the viscosity is close to isotropic in the sense that azimuthal and vertical shear are damped at the same average rates \citep{Torkelssonetal2000,Ogilvie2003,Kingetal2013}. The constraint on the viscosity derived here (eqns \ref{crit} \& \ref{anisocrit}) is consistent with this view.

\section*{acknowledgments}
I thank the referee, Pavel Ivanov, for substantial comments which helped improve the manuscript. I thank Phil Armitage, Andrew King, Steve Lubow and Jim Pringle for useful discussions. I used {\sc splash} \citep{Price2007} for the figure. CJN was supported for this work by NASA through the Einstein Fellowship Program, grant PF2--130098.

\bibliographystyle{mn2e}
\bibliography{nixon}

\appendix

\section{Higher order viscosity coefficients}
\label{coeffs}

In this appendix I describe the relation between the coefficients $\alpha_2$ and $\alpha_3$ derived in \cite{PP1983}, \cite{KP1985} and \cite{Ogilvie1999}.

First, the (linear) derivation by \cite{PP1983} yields an equation for the disc tilt evolution (their 3.17) which is comparable to (\ref{dLdt}) above. Their equation contains the time derivative and advection of the disc tilt with external forcing, equated with the diffusion terms (second and fourth terms of Eq.~\ref{dLdt}). They define the diffusive term through a complex quantity $F$ given by (3.23) in \cite{PP1983}. This quantity $F$ with units of $\nu\Sigma\Omega$ is defined by
\begin{equation}
F = AR\frac{\partial}{\partial R}\left(\frac{U_z}{R\Omega}\right) + B\left(\frac{U_z}{R\Omega}\right)\,,\nonumber
\end{equation}
\begin{equation}
A = \nu\Sigma\Omega + (\nu/H_R^2 - i\Omega)D\,,\nonumber
\end{equation}
\begin{equation}
B = \Sigma\Omega H^2 (\nu/H^2 - i\Omega) + (\nu/H_R^2-i\Omega)E\,,\nonumber
\end{equation}
\begin{equation}
D = \Sigma\Omega H^2[(\nu/H_\phi^2+i\Omega)(i\Omega-\nu/H^2)+3i\nu\Omega/H^2]/X\,,\nonumber
\end{equation}
\begin{equation}
E = -\Sigma\Omega H^2\nu(2i\Omega+\nu/H^2)/(H_\phi^2 X)\nonumber
\end{equation}
and
\begin{equation}
X = i\Omega\nu\left\{\frac{1}{H_R^2} + \frac{1}{H_\phi^2}\right\} + \nu^2/(H_R^2H_\phi^2)\,.
\end{equation}
Here $U_z/(R\Omega)$ is a dimensionless measure of the disc tilt and $H_R$ and $H_\phi$ are defined by (3.21) and (3.22) of \cite{PP1983}. As noted by \cite{PP1983} making the assumption that $\nu$ is independent of the density $\rho$ and that the gas is isothermal yields $H_R=H_\phi=H$. With this and the \cite{SS1973} viscosity $\alpha\Omega = \nu/H^2$, we can simplify the above. In this case
\begin{equation}
X = \Omega^2(\alpha^2 + 2i\alpha)\,,
\end{equation}
which gives 
\begin{equation}
E = -\frac{\Sigma\nu}{\alpha}
\end{equation}
and therefore
\begin{equation}
B = 0\,.
\end{equation}
Whereas
\begin{equation}
D = \Sigma\Omega H^2\frac{-\alpha^2 + 3i\alpha-1}{\alpha(\alpha+2i)}
\end{equation}
and so
\begin{equation}
A = \Sigma\Omega^2 H^2 f(\alpha)\,,
\end{equation}
where 
\begin{equation}
\label{fa}
f(\alpha) = \frac{6i\alpha+2+i/\alpha}{\alpha+2i}\,.
\end{equation}
This is the form of the viscosity coefficient used by \cite{KP1985}. The fact that $B=0$ is a direct consequence of assuming isothermal gas. If the gas is not isothermal then $B\ne 0$ \citep[see][]{Kumar1988}.

In the above equations a complex notation has been employed where the tilt is assumed to remain small (commensurate with the linearisation of the equations). In this case the unit tilt vector $\mathbi{l} = (l_x,l_y,l_z)$ is assumed to have $l_z\approx 1$ and then the equations can be solved for the complex tilt variable $W = l_x+il_y$. It is therefore possible, with a suitable choice of coordinate system, to see that the real and imaginary parts of $f(\alpha)$ correspond to the coefficients of $\nu_2$ and $\nu_3$, which act in the direction of $\partial\mathbi{l}/\partial R$ and $\mathbi{l}\times\partial \mathbi{l}/\partial R$ respectively.

So if we simplify (\ref{fa}) we get real and imaginary parts given by
\begin{equation}
\label{alpha2}
{\rm Re}[f(\alpha)] = \frac{1}{2\alpha}\frac{4(1+7\alpha^2)}{4+\alpha^2}
\end{equation}
and
\begin{equation}
\label{alpha3ish}
{\rm Im}[f(\alpha)]= -\frac{3(1-2\alpha^2)}{4+\alpha^2}\,.
\end{equation}

Now if we note that it is the complex conjugate of $A$ which enters the evolution equation (e.g. equation 2.1 of \citealt{KP1985}) and that the version of the Pringle-Ogilvie equation (\ref{dLdt}) subsumes a factor of a half into $\nu_3$, the relation between these coefficients and those derived by \cite{Ogilvie1999} becomes clear. (\ref{alpha2}) gives the relevant value of $\alpha_2$ and $-$(\ref{alpha3ish})$/2$ gives the value of $\alpha_3$, as given by equation~137 of \cite{Ogilvie1999} and the relations $\alpha_2=2Q_2$ and $\alpha_3=Q_3$. Thus the $\alpha$ dependence derived by \cite{PP1983} and \cite{Ogilvie1999} are identical. \cite{Ogilvie1999} extends these coefficients to the nonlinear warp regime \citep[see also][]{Ogilvie2000,OL2013b,OL2013a}, whereas \cite{PP1983} is valid for the linear regime.

For small $\alpha \ll 1$ we can see that these recover the classical values of $\alpha_2$ and $\alpha_3$ relevant for the small warp and small $\alpha$ case. By ignoring terms of order $\alpha^2$ we have
\begin{equation}
\alpha_2 = \frac{1}{2\alpha}
\end{equation} 
and
\begin{equation}
\alpha_3 = \frac{3}{8}\,.
\end{equation}

\section{Effects of non-Keplerian potential}
\label{nonkep}
The disc in Her X-1 orbits in the potential of a binary system. Therefore the disc rotation, particularly in the outer regions is not strictly Keplerian. The warped disc evolution equation (\ref{dLdt}) accounts for an arbitrary rotation law $\Omega$ \citep{Pringle1992}, and the radiation warping criterion is derived from this general form \citep{Pringle1996}. The main effect of non-Keplerian rotation is to alter the values of the viscosity coefficients in (\ref{dLdt}). The instability of a disc to radiation warping requires an initially planar disc (with a small perturbation) to be linearly unstable (see e.g. Section 3 of \citealt{Pringle1996}) and therefore it is appropriate to explore the linear warp viscosity coefficients.

The leading order corrections to the coefficients, in the limit of small amplitude warps, is given by Equation~A39 in \cite{Ogilvie1999}. This is
\begin{equation}
Q_{40} = \frac{1}{2}\left[\frac{i-2\alpha + (7-\tilde{\kappa}^2)i\alpha^2}{(1-\tilde{\kappa}^2)+2i\alpha-\alpha^2}\right]\,,
\end{equation}
where $\tilde{\kappa}=\kappa/\Omega$ is the dimensionless epicyclic frequency and $Q_{40}=Q_{20}+iQ_{30}$ is the viscosity coefficients evaluated at linear warp amplitudes (e.g. $Q_i = Q_{i0} + \left|\psi\right|^2 Q_{i2} + \mathcal{O}(\left|\psi\right|^4)$). By suitable rearrangement (set $\tilde{\kappa}=1$, multiply by $-1/-1$, times by 2 and then take the complex conjugate) this recovers (\ref{fa}).

Now including these extra non-Keplerian terms, as above we have $\alpha_2 = 2{\rm Re}[Q_{40}]$. So
\begin{equation}
\alpha_2 = {\rm Re}\left[\frac{i-2\alpha + (7-\tilde{\kappa}^2)i\alpha^2}{(1-\tilde{\kappa}^2)+2i\alpha-\alpha^2}\right]\,.
\end{equation}
Simplifying and taking the real part gives
\begin{equation}
\label{alpha2nonkep}
\alpha_2 = \frac{2\alpha\tilde{\kappa}^2 + 2(8-\tilde{\kappa}^2)\alpha^3}{(1-\tilde{\kappa}^2 - \alpha^2)^2+4\alpha^2}\,.
\end{equation}
With $\tilde{\kappa}=1$ this recovers (\ref{alpha2}).

To evaluate this expression, for comparison with (\ref{alpha2}) we must evaluate $\tilde{\kappa}^2$. The companion star induces perturbations to the disc orbits which deviate from Keplerian rotation. This effect is strongest at large disc radii closest to the companion. So for a conservative estimate of the non-Keplerian effects we can focus on the disc outer edge at $R=0.3R_{\rm b}$. The equations for $\kappa^2$ and $\Omega^2$ in this potential are given by \citep[e.g.][]{LO2000}
\begin{equation}
\Omega^2 = \frac{GM_1}{R^3} + \frac{GM_2}{2R_{\rm b}^2R}\left[\frac{R}{R_{\rm b}}b_{3/2}^{(0)}\left(\frac{R}{R_{\rm b}}\right) - b_{3/2}^{(1)}\left(\frac{R}{R_{\rm b}}\right)\right]
\end{equation}
and
\begin{equation}
\kappa^2 = \frac{GM_1}{R^3} + \frac{GM_2}{2R_{\rm b}^2R}\left[\frac{R}{R_{\rm b}}b_{3/2}^{(0)}\left(\frac{R}{R_{\rm b}}\right) - 2b_{3/2}^{(1)}\left(\frac{R}{R_{\rm b}}\right)\right]\,,
\end{equation}
where
\begin{equation}
b_{\gamma}^{(m)}(x) = \frac{2}{\pi}\int_0^\pi \cos(m\phi)(1+x^2-2x\cos\phi)^{-\gamma} {\rm d}\phi
\end{equation}
is the Laplace coefficient.

\begin{figure}
  \center{\includegraphics[width=\columnwidth]{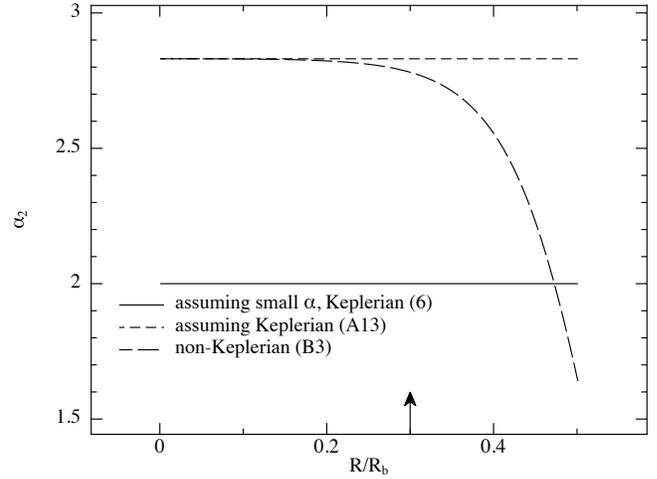}}
  \caption{Comparison of the three different calculations of the viscosity coefficient $\alpha_2$ for $\alpha=0.25$ and an equal-mass binary. The solid line is the standard (small $\alpha$ and Keplerian) value given by (\ref{visccoef}), the short-dashed line includes the higher order correction in $\alpha$ given by (\ref{alpha2}), and the long-dashed line includes the non-Keplerian terms from the binary potential given by (\ref{alpha2nonkep}). The assumed outer radius of the disc is shown by the arrow on the $x$-axis at $R=0.3R_{\rm b}$.}
  \label{alpha2s}
\end{figure}

Numerical evaluation at $R=0.3R_{\rm b}$, for an equal mass binary, yields $\tilde{\kappa}^2=0.95$. Evaluating (\ref{alpha2nonkep}) with this value gives $\alpha_2 = 2.78$. This is close to the Keplerian value of $2.8$ given by (\ref{alpha2}). The different calculations of $\alpha_2$ are shown in Fig.~\ref{alpha2s} This shows that the higher order $\alpha$ terms are more significant than the non-Keplerian terms for this system in determining $\alpha_2$. With both being close enough to the standard value that its use in Section~\ref{constraint} above is justified. However, these additional calculations should be taken into account in more extreme systems, such as the inner parts of discs around Kerr black holes.

\section{Higher order viscosity coefficients in the anisotropic case}
\label{anisocoeffs}
In Appendix~\ref{coeffs} I discuss the higher order corrections to the viscosity coefficients assuming an isotropic viscosity. In this appendix I detail the same derivation, but retaining the anisotropic generality.

Following \cite{PP1983} and noting that the $R-\phi$ stress coefficient is $\alpha_{\rm h}$, the $R-z$ stress coefficient is $\alpha_{\rm v}$ and the $\phi-z$ stress coefficient is (previously unintroduced) $\alpha_{\rm t}$ (corresponding to $\alpha_3$ above), it can be seen that their (3.21) corresponds to $\alpha_{\rm v}$ and (3.22) corresponds to $\alpha_{\rm t}$. From this, and inspection of their (3.3) and (3.19), then (cf. Appendix~\ref{coeffs})
\begin{equation}
A = \nu_{\rm v}\Sigma\Omega + (\nu_{\rm v}/H_R^2 - i\Omega)D
\end{equation}
where I have adopted the notation $\nu_i = \alpha_i H^2 \Omega$. Also,
\begin{equation}
D = \Sigma\Omega H^2[(\nu_{\rm t}/H_\phi^2+i\Omega)(i\Omega-\nu_{\rm h}/H^2)+3i\nu_{\rm h}\Omega/H^2]/X
\end{equation}
and
\begin{equation}
X = i\Omega\left\{\frac{\nu_{\rm v}}{H_R^2} + \frac{\nu_{\rm t}}{H_\phi^2}\right\} + \nu_{\rm v}\nu_{\rm t}/(H_R^2H_\phi^2)\,.
\end{equation}

Assuming the same simplifications as before, this simplifies to
\begin{equation}
X = \alpha_{\rm v}\alpha_{\rm t}\Omega^2 + i\Omega^2(\alpha_{\rm v} + \alpha_{\rm t})\,,
\end{equation}
\begin{equation}
D = \Sigma\Omega H^2 \frac{\left[(\alpha_{\rm t}+i)(i-\alpha_{\rm h}) + 3i\alpha_{\rm h}\right]}{\alpha_{\rm v}\alpha_{\rm t} + i(\alpha_{\rm v}+\alpha_{\rm t})}
\end{equation}
and
\begin{equation}
A = \Sigma\Omega^2 H^2\left[\alpha_{\rm v} + \frac{(\alpha_{\rm v}-i)([\alpha_{\rm t}+i][i-\alpha_{\rm h}] + 3i\alpha_{\rm h})}{\alpha_{\rm v}\alpha_{\rm t} + i(\alpha_{\rm v}+\alpha_{\rm t})}\right]\,.
\end{equation}

This gives
\begin{equation}
f(\alpha) = \alpha_{\rm v} + \frac{(\alpha_{\rm v}-i)([\alpha_{\rm t}+i][i-\alpha_{\rm h}] + 3i\alpha_{\rm h})}{\alpha_{\rm v}\alpha_{\rm t} + i(\alpha_{\rm v}+\alpha_{\rm t})}
\end{equation}
and therefore the real part gives
\begin{eqnarray}
  \alpha_2 & = & (\alpha_{\rm v}^3\alpha_{\rm t}^2 - \alpha_{\rm h}\alpha_{\rm v}^2\alpha_{\rm t}^2 + \alpha_{\rm v}^3 + 3\alpha_{\rm v}\alpha_{\rm t}^2 + \alpha_{\rm h}\alpha_{\rm t}^2 + 2\alpha_{\rm v}^2\alpha_{\rm h} \\ \nonumber & &
  +{}~2\alpha_{\rm v}^2\alpha_{\rm t} + 5\alpha_{\rm h}\alpha_{\rm v}\alpha_{\rm t} + \alpha_{\rm v} + \alpha_{\rm t})/(\alpha_{\rm v}^2 \alpha_{\rm t}^2 + \alpha_{\rm v}^2 + \alpha_{\rm t}^2 + 2\alpha_{\rm v}\alpha_{\rm t})\,.
\end{eqnarray}

This can be understood by taking various limits. By setting $\alpha_{\rm h} = \alpha_{\rm v} = \alpha_{\rm t} = \alpha$ it can be seen that this recovers (\ref{alpha2}). If we assume all of the coefficients are small (dropping terms of order $\alpha_i^3$) and $\alpha_{\rm t}\approx\alpha_{\rm v}$, this becomes
\begin{equation}
\alpha_2 \approx \frac{1}{2\alpha_{\rm v}}\,.
\end{equation}
Similarly, assuming all $\alpha_i \sim 1$, then $\alpha_2 \approx 3.2$.

Note that in this anisotropic case, $B \ne 0$. This means that the diffusion of tilt governed by the $\nu_2$ term in (\ref{dLdt}) becomes an advection-diffusion term (see \citealt{Kumar1988}) and this must be included for a complete picture.

\label{lastpage}
\end{document}